\documentclass[%
 reprint,
 onecolumn,
 amsmath,amssymb,
 pra,
]{revtex4-2}

\usepackage{graphicx}
\usepackage{dcolumn}
\usepackage{bm}

\newcommand{\ket}[1]{|#1\rangle}
\newcommand{\bra}[1]{\langle #1|}

\graphicspath{{./pics/}}

\usepackage[figurename=FIG., justification=RaggedRight]{caption}
\usepackage{subcaption} 
\usepackage{physics}

\usepackage[titletoc]{appendix}

\begin{document}

\title{Joint eavesdropping on the BB84 decoy state protocol with an arbitrary passive light-source side channel}

\author{D. V. Babukhin$^{1,2}$ and D.V. Sych$^{2,3}$}

\affiliation{$^1$QRate LLC, Novaya av. 100, Moscow 121353, Russia}
\affiliation{$^2$Department of Mathematical Methods for Quantum Technologies, Steklov Mathematical Institute of Russian
Academy of Sciences, Gubkina str. 8, Moscow 119991, Russia}
\affiliation{$^3$P.N. Lebedev Physical Institute, Russian Academy of Sciences, 53 Leninskiy Prospekt, Moscow 119991, Russia}

\begin{abstract}
Passive light-source side channel in quantum key distribution (QKD) makes the quantum signals more distinguishable thus provides additional information about the quantum signal to an eavesdropper.
The explicit eavesdropping strategies aimed at the passive side channel known to date were limited to the separate measurement of the passive side channel in addition to the operational degree of freedom.
Here we show how to account for the joint eavesdropping on both operational degree of freedom and the passive side channel of the generic form.
In particular, we use the optimal phase-covariant cloning of the signal photon state, which is the most effective attack on the BB84 protocol without side channels, followed by a joint collective measurement of the side channel and the operational degree of freedom. To estimate QKD security under this attack, we develop an ``effective error'' method and show its applicability to the BB84 decoy-state protocol. 

\end{abstract}

\maketitle
\date{\today }


\section{Introduction}

Quantum key distribution (QKD) provides theoretically secure communication between legitimate sides \cite{LoChau1999, Shor2000, gisin2002}.
In practice, various deviations of real devices from theoretical models lead to overestimation of security of a real QKD setup \cite{Scarani2014}. Whenever a particular device behaves differently from its theoretical model, the eavesdropper (Eve) can use this difference to construct a more efficient attack on the QKD protocol \cite{Nauerth2009, Pereira2019}. This additional eavesdropping option reduces QKD security and forms a so-called informational side channel that can completely compromise the QKD protocol \cite{Makarov2017, Huang2018}. 
The goal of practical quantum communication is to carefully analyse and estimate opportunities of eavesdropping beyond those allowed by an ideal theoretical model of QKD \cite{Diamanti2016, Xu2020, Jain2016}.

Among all opportunities to attack a given experimental realization of QKD, there are attacks on the quantum source (Alice) and attacks on the receiver (Bob). Historically, the receiver in QKD is more prone for hacking \cite{Jain2016}. Fortunately, a measurement-device-independent QKD protocol allows closing every possible quantum hacking on the receiver side \cite{MDI} and still obtain practically-valuable secure key rates over long distances. The device-independent QKD \cite{Acin2007} excludes side channels of the source and the receiver simultaneously, but at the cost of complicated experimental realization and low secret key rate at practically reasonable distances. The other way to deal with light-source side channels is to analyse Eve's opportunities to gain information from them. 
The most general analysis of device imperfections up to date is GLLP \cite{Gottesman2002}, which allows to estimate information leakage from the general principles. This approach provides pessimistically-low secure distances, thus there is a need for searching explicit attacks with more practical security characteristics.
Active-probing-based side channels (so called ``Trojan horse'' attack) are widely analysed in the literature and explicit attacks satisfying the GLLP bound are partially provided \cite{Lucamarini2015}. At the same time, analysis of passive light-source side channels (i.e. distinguishability of photon source in non-operational degrees of freedom \cite{Nauerth2009}) are not so developed. There is a huge gap between the lower and upper bounds on the secret key, derived from the general principles \cite{Gottesman2002} and from explicit attacks on QKD with passive side channels, respectively \cite{Babukhin2020, Babukhin2021, Sych2021}. The gap can be potentially reduced by tightening the bounds. 

In this paper, we tighten the upper bound on the secret key and consider a joint eavesdropping strategy on the signal and passive side channel states of the general form in the BB84 decoy-state protocol.
This strategy consists of a phase-covariant cloning of the operational degree of freedom and a joint measurement of both operational and a non-operational degrees of freedom. 
We introduce passive source side channel model as an additional degree of freedom, which has dimension of the QKD protocol alphabet and allow accounting for arbitrary distinguishability of signal photons.
To estimate efficiency of this attack on the decoy-state protocol, we propose a method to calculate secret key rates in QKD protocols with explicit attacks on the signal photon state and on the passive source side channel. 
Namely, we show how to reinterpret information flows among Bob and Eve, and calculate an effective error rate on Bob's side, which allows incorporating theoretically calculated error into the decoy state protocol. 

This paper is organized as follows. 
In Sec. II.A we introduce background for the single-photon BB84 protocol and show a model of a light-source side channel. 
In Sec. II.B we discuss an application of the Hong-Ou-Mandel (HOM) interference to estimation of information leakage through the light source side channel.
In Sec. II.C we discuss eavesdropping on the operational degree of freedom.
In Sec. II.D we introduce an ``effective error'' method, which we further use to estimate security of the decoy-state BB84 protocol with photons' distinguishability.
In Sec. II.E we provide calculation results and connect it to the state-of-the-art photon sources.
%
Description of the BB84 protocol with decoy states and how it incorporates the effective error into security analysis is provided in Appendix.


\section{BB84 with source side channel and effective error rate calculation}

\subsection{Side channel model}

In the BB84 protocol, legitimate sides exchange bits encoded in quantum states of signal photons. Alice randomly chooses a bit value (0 or 1) and a basis to encode the bit in (X or Y). Then she sends the encoded state into a quantum channel, which is open for eavesdropping. The eavesdropper attacks the photon and thus introduces errors in communication, which alert the legitimate sides about eavesdropping. 
Without photon distinguishability, photon states in BB84 protocol are
\begin{equation}
    \label{ensembleBB84}
    \biggl{\{} \frac{1}{4}: 
        \ket{0_{x}},\text{  }
       \frac{1}{4}:
       \ket{1_{x}},\text{  }
       \frac{1}{4}:
       \ket{0_{y}},\text{  }
       \frac{1}{4}:
       \ket{1_{y}}
      \biggl{\}}.
\end{equation}
If photons are distinguishable other than in signal degree of freedom, we need to incorporate this distinguishability into the model through an additional degree of freedom. This degree of freedom is non-operational for legitimate users, but it is visible to Eve, who can extract additional information.
For example, if Alice and Bob use photons polarization to encode secret bits, differences in other photon degrees of freedom (e.g., spatial modulation, frequencies) provide additional information to Eve. Thus, there is a side channel of information, which leaks information to Eve.
This transforms BB84 protocol states to the form
\begin{equation}
    \label{ensembleBB84sidechannel}
    \biggl{\{} \frac{1}{4}: 
        \ket{0_{x}}\otimes\ket{0^{\Delta}_{X}},\text{  }
       \frac{1}{4}:
       \ket{1_{x}}\otimes\ket{1^{\Delta}_{X}},\text{  }
       \frac{1}{4}:
       \ket{0_{y}}\otimes\ket{0^{\Delta}_{Y}},\text{  }
       \frac{1}{4}:
       \ket{1_{y}}\otimes\ket{1^{\Delta}_{Y}},\text{  }
      \biggl{\}},
\end{equation}
where states $\ket{0^{\Delta}_{X}}$, $\ket{1^{\Delta}_{X}}$, $\ket{0^{\Delta}_{Y}}$, $\ket{1^{\Delta}_{Y}}$ are nonorthogonal states of additional degree of freedom, which models the photons' distinguishability side channel. 
Here we introduce passive source side channel model, which has the most general form without any symmetry constraints (unlike in the previous studies \cite{Babukhin2022}) and allow accounting for any kind of distingusihability in signal photon non-operational properties.
The number of lasers, which compose the QKD source and produce signal photons, dictates the number of  states which compose a basis of side channel state space. For example, if a QKD source has four different lasers, then, in general, all four states of side channel degree of freedom have different non-unity scalar product with other side channel states. This case models the situation, when non-operational degrees of freedom of photons, produced with different lasers, are not completely coincident due to imprecise lasers calibration or some unnoticed device flaw. 

The described model allows estimating influence of side channel on the security of QKD in the whole range of non-operational photon distinguishability. The case of indistinguishable side-channel states corresponds to the protocol without excess information leakage, and the case of orthogonal side-channel states corresponds to a protocol, completely compromised through side channels. 

\subsection{Hong-Ou-Mandel visibility as a measure of side channel leakage}

One can test physical difference of photons using Hong-Ou-Mandel interference \cite{Hong1987,Branczyk2017}. The HOM interference is a fourth-order interference of photons, which makes two physically indistinguishable photons, which are incident on a beam splitter, to exit this beam splitter pairwise in one or another arm. If photodetectors are placed in front of each arm, there will be no simultaneous counts of both detectors, i.e., no coincident counts. Contrary, if photons are physically distinct (e.g., some of their modes have different states), they can leave beam splitter in different arms and produce coincident counts of two photodetectors. 

This effect can be used to estimate passive source side channels in QKD. In particular, the more distinguishability of photons in non-operational degrees of freedom, the more information can potentially leak to Eve through the side channel.  
Alice can use HOM interference to estimate photons distinguishability in all non-operational degrees of freedom at once, and thus to estimate information leakage \cite{Duplinskii2019}. She needs to bring two photons from difference source lasers into one operational degree of freedom (e.g., polarization) and send them to two arms of a balanced beam splitter to measure interference visibility. If photons' states are described with density matrices $\rho_{1}$ and $\rho_{2}$, the HOM interference visibility is equal to
 \begin{equation}
     V(\rho_{1}, \rho_{2}) = Tr[\rho_{1} \rho_{2}] = \frac{N_{max} - N_{min}}{N_{max}},
 \end{equation}
where $N_{min}$ and $N_{max}$ are minimum and maximum values of coincidence counts.
If visibility is maximal (1 for single photons), then these photons are not distinct in any degrees of freedom. If visibility is minimal (0 for single photons), then photons are completely distinct in some degree of freedom and can be discriminated through a proper quantum state measurement. The visibility value allows to estimate a basis imbalance parameter, which characterizes information leakage through the side channel \cite{Duplinskii2019}:
\begin{eqnarray}
    \label{DeltaDuplinskiy}
   \Delta \leq \frac{1}{2}
   \biggl(
       1 - \cos\biggl( 2 \arccos{\frac{1 + e^{\mu(\sqrt{2V}-1)}}{2}} + \arccos{e^{\mu(\sqrt{2V}-1)}} \biggl)
   \biggl)
\end{eqnarray}
where $\mu$ is a signal pulse intensity and $V$ is a visibility value. More discussion about parameter $\Delta$ see in Appendix \ref{AppendixA3}. The formula (\ref{DeltaDuplinskiy}) allows connecting theoretical parameter $\Delta$ with the HOM visibility of practical QKD sources. We will use this connection in the following sections.

\subsection{Eavesdropping on the signal degree of freedom}

Here we describe a unitary attack on the signal degree of freedom, which is a standard action of Eve in the process of QKD communication. Because side channel gives Eve only partial information about the secret key, Eve combines attacking side channel with attack on the signal photon. 
The most effective attack on photons in the BB84 (a so called collective attack) gives equal information flows from Alice to Bob and from Alice to Eve, when Bob has a communication error equal to $11\%$. This attack can be implemented with an optimal phase-covariant cloning machine \cite{Bruss2000}.
For signal states of the BB84 from the $XY$ plane of the Bloch sphere, the action of this cloning machine is
\begin{eqnarray}
    U\ket{\psi(\phi)}_{B}\ket{0_{z}}_{E}\ket{0_{z}}_{E^{'}} = \frac{1}{2}(\ket{0_{z}}_{B}\ket{0_{z}}_{E}\ket{0_{z}}_{E^{'}} +
    \notag\\
    + \cos\eta\ket{0_{z}}_{B}\ket{1_{z}}_{E}\ket{1_{z}}_{E^{'}}
    + \sin\eta\ket{1_{z}}_{B}\ket{0_{z}}_{E}\ket{1_{z}}_{E^{'}} \pm
    \notag\\
    \pm \cos\eta\ket{1_{z}}_{B}\ket{0_{z}}_{E}\ket{0_{z}}_{E^{'}} \pm 
    \sin\eta\ket{0_{z}}_{B}\ket{1_{z}}_{E}\ket{0_{z}}_{E^{'}} \pm 
    \notag\\
    \pm \ket{1_{z}}_{B}\ket{1_{z}}_{E}\ket{1_{z}}_{E^{'}}),
\end{eqnarray}
where $\eta$ is a cloning parameter. When $\eta = 0$, the cloner unitary does nothing, and when $\eta = \pi/2$, Eve has Bob's state in her space and Bob's qubit becomes maximally mixed because of entanglement with Eve's ancillary qubit.

This attack leads to a critical error value $Q_{c} \approx 0.11$ for the standard BB84 protocol with photons, indistinguishable in non-operational degrees of freedom. 
The attack on the signal degree of freedom consists of Eve doing a quantum cloning transform on the Alice signal photon, thus correlating the photon with Eve's ancillary system, and waiting for basis exchange. Eve's ancillary system is considered a quantum memory, which can store a quantum state for an infinitely long time. After Alice and Bob exchange their basis choices on all bit positions, Eve makes a full register measurement and obtains a binary string, which is correlated with distributed bits sequence between Alice and Bob. 

The amount of information, which Eve obtains during the collective attack on the QKD protocol is characterized with a Holevo value. This value upper-bounds possible mutual information between Alice and Eve. In the BB84 protocol with balanced basis choice ($\frac{1}{2}$ of times Alice chooses basis $X$ and $\frac{1}{2}$ of times Alice chooses basis $Y$), the Holevo value is calculated as follows
\begin{equation}
    \chi = S(\frac{1}{2}\rho_{0,X} + \frac{1}{2}\rho_{1,X}) - \frac{1}{2}S(\rho_{0,X}) - \frac{1}{2}S(\rho_{1,X}),
\end{equation}
where $S$ denote a von Neuman entropy
\begin{equation}
    S(\rho) = -Tr(\rho\log(\rho)),
\end{equation}
where $\rho_{0,X}$ and $\rho_{1,X}$ are Eve states, obtained with cloning states in the quantum channel between Alice and Bob.

\subsection{Effective error in QKD with a side channel}

Side channels increase Eve's information about the secret bit, because Eve has more resources to distinguish collected quantum states.
It leads to increase of information leaking to Eve, and it formally leads to increase of Eve's mutual information bounded with a Holevo value
\begin{equation}
    \chi < \chi^{\Delta},
\end{equation}
where 
\begin{equation}
    \chi^{\Delta} = 
    S
    \biggl(
    \frac{1}{2}\rho_{0,X}\otimes\ket{0^{\Delta}_{X}}\bra{0^{\Delta}_{X}} + \frac{1}{2}\rho_{1,X}\otimes\ket{1^{\Delta}_{X}}\bra{1^{\Delta}_{X}}
    \biggl) - 
    \frac{1}{2}S
    \biggl(
    \rho_{0,X}\otimes\ket{0^{\Delta}_{X}}\bra{0^{\Delta}_{X}}
    \biggl) - 
    \frac{1}{2}
    S
    \biggl(
    \rho_{1,X}\otimes\ket{1^{\Delta}_{X}}\bra{1^{\Delta}_{X}}
    \biggl)
\end{equation}
is a Holevo value of Eve's attack on the protocol with side channels.
This value is equal to $\chi$, when side channel states coincide ($\bra{1^{\Delta}_{X}}\ket{0^{\Delta}_{X}} = 1$), which means there is no information leakage through side channel.

We can look at the increase of information leakage at a different perspective. With more information, Eve has an information channel from Alice with less error, while Bob's information channel does not change. The secret key rate is calculated through subtraction of channels capacities between Alice and Bob, and Alice and Eve. Formally, we can write this equation twofold:
\begin{equation}
    \label{Rdelta}
    R^{\Delta} = 1 - h_{2}(Q_{Bob}) - \chi^{\Delta} = 1 - h_{2}(Q_{Bob}^{\Delta}) - \chi.
\end{equation}
Here, $Q_{Bob}$ is an error on Bob's side in the QKD protocol, which occurs due to the unitary eavesdropping of the signal photon state. This error is not influenced with side channel attack, and using this quantity for further protocol analysis will give an overestimated security. Instead, we can consider the secret key rate as written on the right hand side on the equation (\ref{Rdelta}): here Eve obtains no information from side channel, but Bob obtains less information due to a larger ("effective") error rate $Q_{Bob}^{\Delta}$, which accounts for information leakage through side channels.  

This equivalence allows us to calculate the effective error of Bob with use of explicit attacks on the protocol, such as unitary attacks on the signal, explicit attacks on the side channel states as well as the model of side channel. The result of combining these parts of eavesdropping is a single value - effective Bob error - which can be used for further security estimation of the protocol. 

Calculation of the effective error consists of the following steps:
\begin{enumerate}
    \item Choose a model of a source side channel in the protocol;
    \item Choose an attack on the side channel states (i.e., how Eve measures side channel states and how she uses this information)
    \item Choose a unitary attack to eavesdrop the signal photon;
    \item Calculate effective error on Bob's side $Q_{Bob}^{\Delta}$ with equation (\ref{Rdelta});
    \item Use the effective error $Q_{Bob}^{\Delta}$ for further security analysis.
\end{enumerate}

In the following we provide a concrete example of using this effective error approach.

\subsection{Results}

Here we apply the effective error approach to calculation of the explicit attack strategy on the BB84 protocol with decoy states. Details of the decoy-state method are provided in Appendix \ref{AppendixA1}. We show how the effective error corrects detection error on the Bob's side in Appendix \ref{AppendixA2}. We show how our side channel model we use here performs in GLLP approach \cite{gottesman2002security} in Appendix \ref{AppendixA3}. 
In Fig. \ref{fig: Figure1} we provide secret key rates for the case, when Eve does no eavesdropping on the signal photon state and only attacks a side channel degree of freedom. In our calculation, this case corresponds to the optimal phase-covariant cloning with $\eta = 0$.  In Fig. \ref{fig: Figure2} we provide secret key rates for the case, when Eve does eavesdropping both on the signal photon and on the side channel state. We compare our approach with the GLLP secret key estimate.

In simulations of the decoy state protocol we used fiber attenuation $\alpha = 0.2$, dark count probability $Y_0 = 10^{-5}$, an average number of photons per pulse $\mu = 0.5$, optical error rate $1\%$ and error correction efficiency $f = 1.0$.

%
%

\begin{figure}[h!]
	\includegraphics[width=0.75\linewidth]{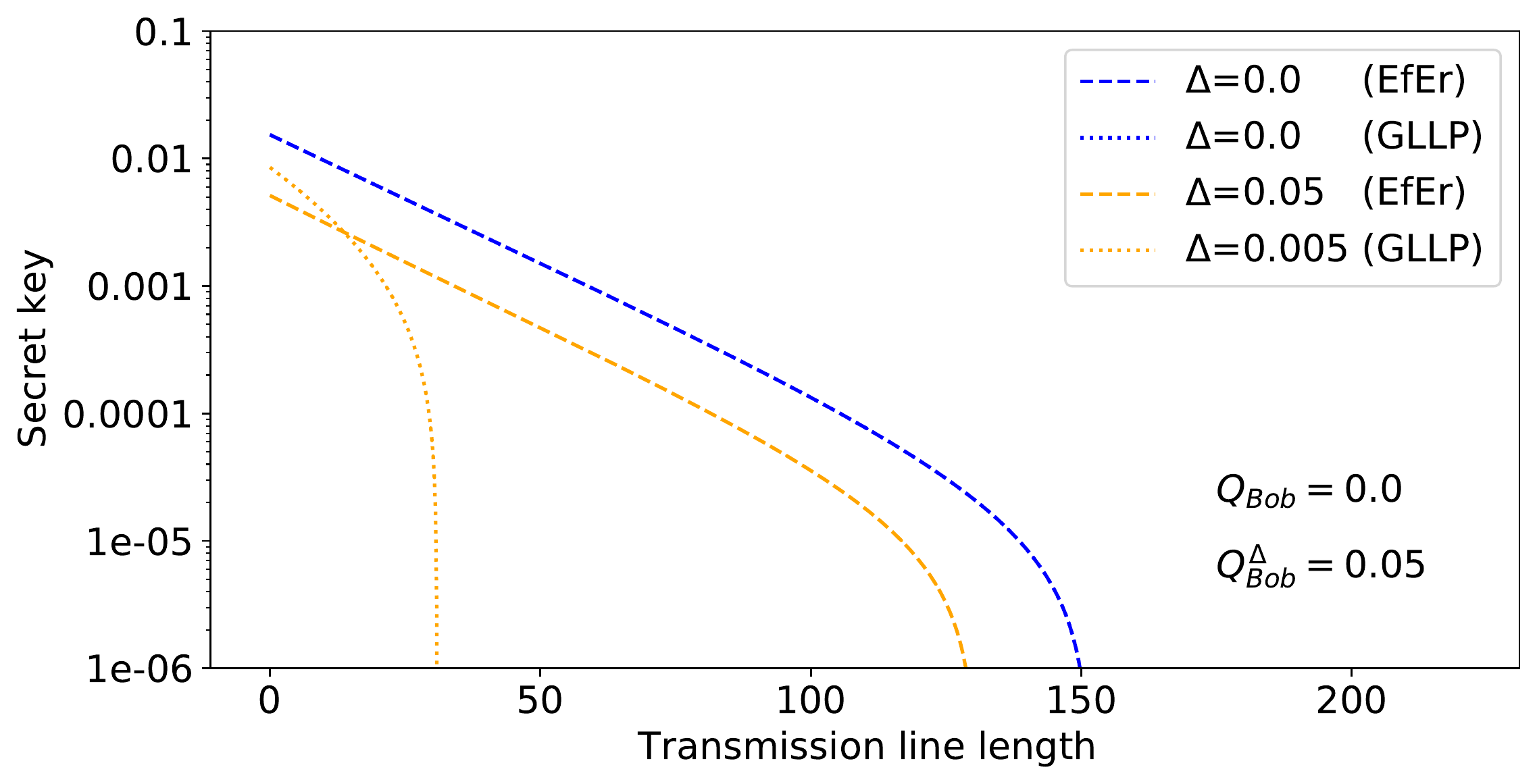}
	\caption{Secret key rates for the BB84 decoy state protocol with a passive source side channel. Here Eve does no eavesdropping on the signal photon state. The imbalance value $\Delta$ indicates amount of information leakage through the side channel, ``EfEr'' label stands for ``effective error'' method.}
	\label{fig: Figure1}
\end{figure}

\begin{figure}[h!]
	\includegraphics[width=0.75\linewidth]{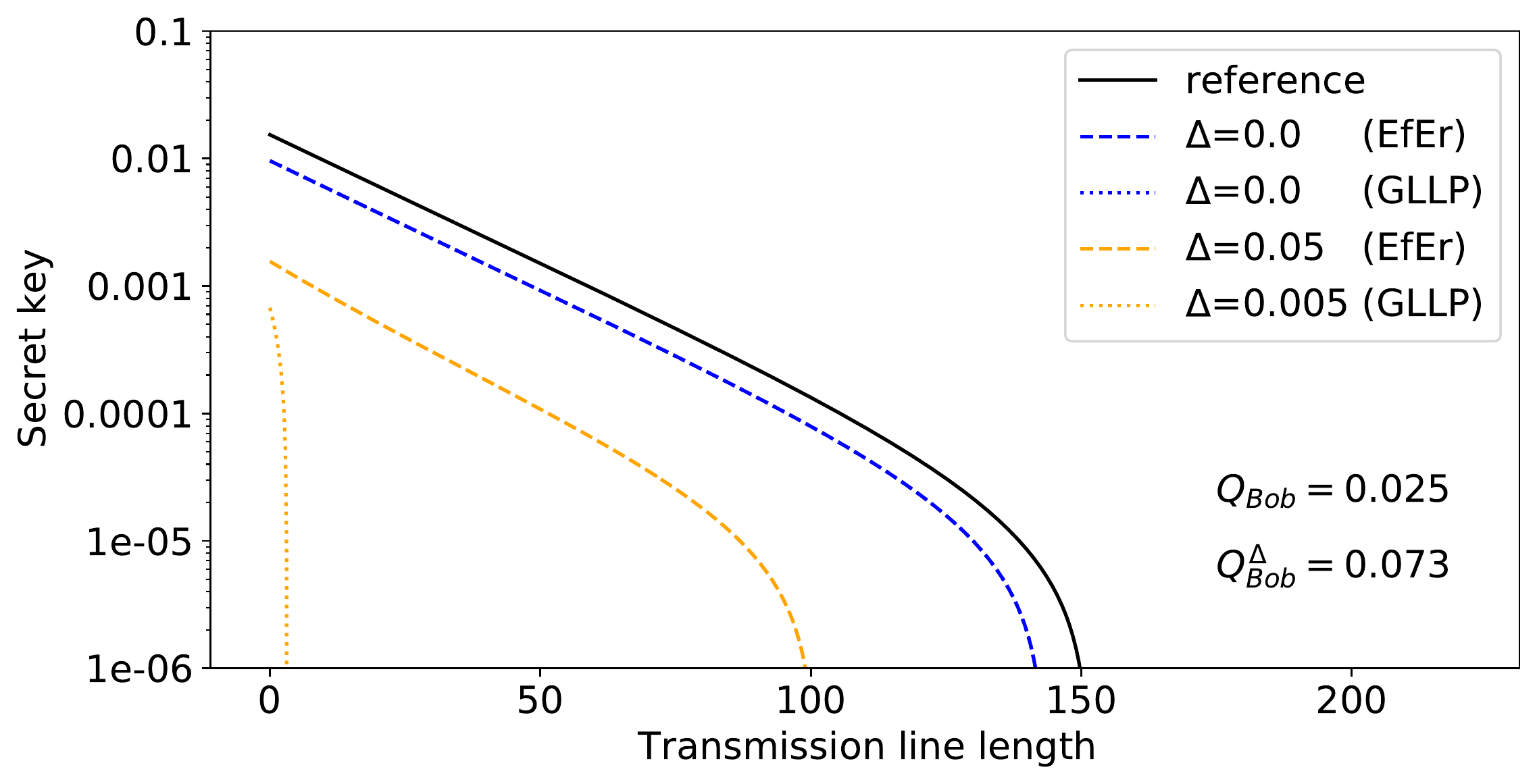}
	\caption{Secret key rates for the BB84 decoy state protocol with a passive source side channel. Here Eve does eavesdropping (phase-covariant cloning) the signal photon state. The imbalance value $\Delta$ indicates amount of information leakage through the side channel, ``EfEr'' label stands for ``effective error'' method. The reference curve indicates the protocol with no eavesdropping on signal and side channel states.}
	\label{fig: Figure2}
\end{figure}

\begin{figure}
    \begin{minipage}[b]{0.99\textwidth}
        \centering
        \includegraphics[width=0.75\linewidth]{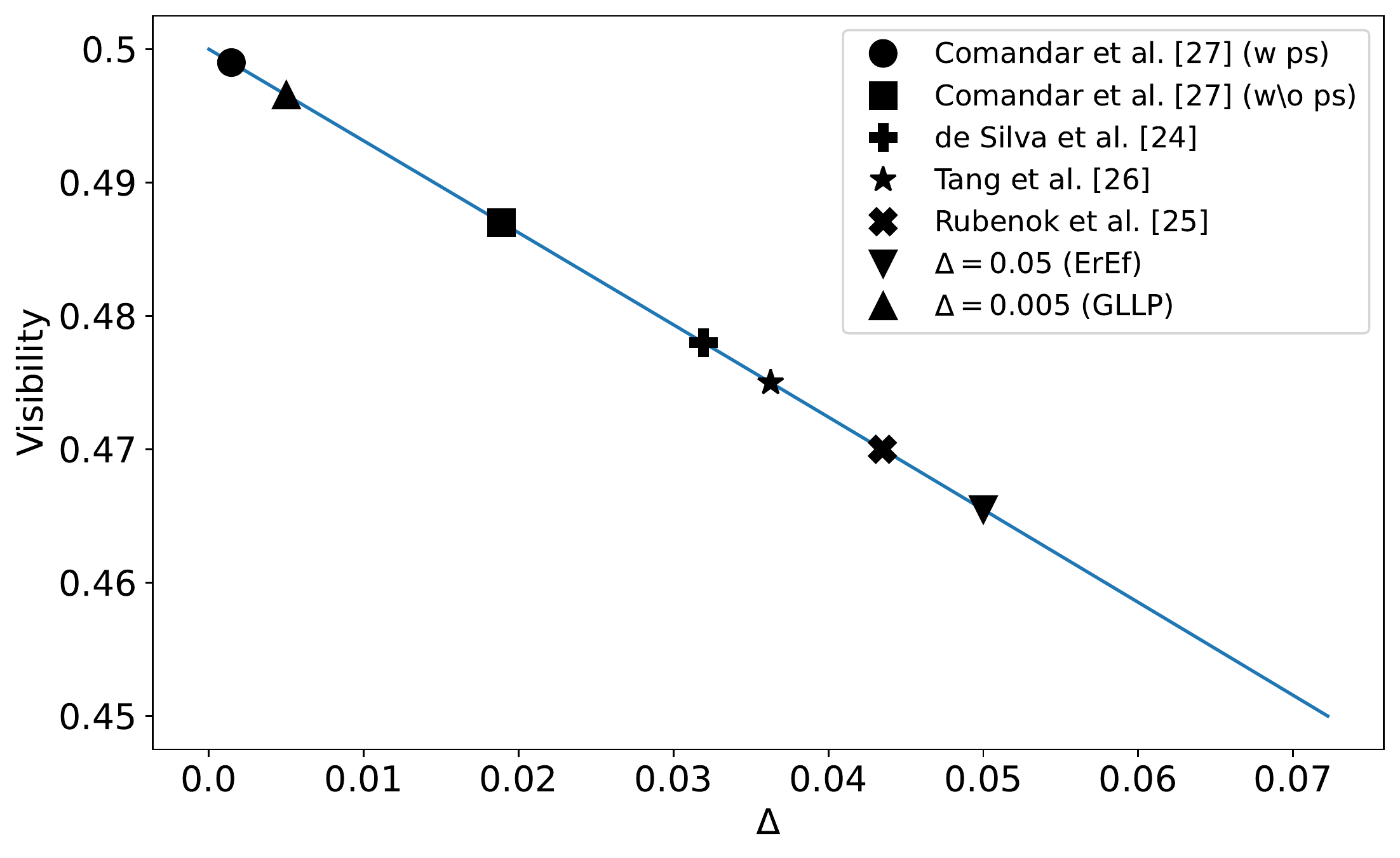}
        \caption{Visibility vs. basis imbalance parameter $\Delta$, derived from Eq.~\ref{DeltaDuplinskiy}. Values of $\Delta$ correspond to the state-of-the-art photon sources \cite{HOMvis1,HOMvis2,HOMvis3,HOMvis4}. Values of $\Delta$, which we use in our simulations, are also provided here for reference (up and down triangles).}
        \label{fig:visibility}
    \end{minipage}
\end{figure}

The results demonstrate an application example of the effective error approach to the analysis of the explicit attack on the protocol with a passive source side channel.
The effective error is above zero in the absence of eavesdropping the signal degree of freedom and there is a corresponding decrease of secret key generation rate (see Fig.~\ref{fig: Figure1}). This result is expected since there is a leakage of information to Eve through the side channel.
If the signal degree of freedom is also eavesdropped, the effective error characterizes the overall leakage of information (see Fig.~\ref{fig: Figure2}).

In our calculation we used an optimal phase-covariant cloning, which is the most efficient attack on the BB84 protocol without side channels \cite{Brus2000}, along with a joint collective measurement of a "signal + side channel" system. 
The joint collective measurement of all degrees of freedom (both operational and non-operational) is a stronger eavesdropping strategy of using passive source side channel, compared to those covered in the previous studies \cite{Babukhin2022}. 
This attack strategy uses information from side channel nonadaptively - Eve obtains additional information and does not actively uses this information to change a signal photon state (e.g., make states of the protocol alphabet more orthogonal or block some photons with respect to a particular criterion). 
We calculated secret key rates for values of $\Delta$, which correspond to visibility values of the state-of-the-art photon sources (see Fig.~\ref{fig:visibility}). 
Thus, secret key rates we calculated provide a practical upper bound of the key generation rate for a considered attack strategy.


\section{Conclusion}

We analyzed joint eavesdropping on the operational degree of freedom and a passive side channel in the BB84 decoy state protocol with a photon distinguishability side channel.
We used an effective error approach to estimate the security of QKD protocols with side channels. 
Our results allowed us to tighten the upper bound on the secret key rate of the BB84 decoy state protocol with a passive source side channel. Since the BB84 with decoy states comprises a backbone of the most practically-available QKD systems, our result can potentially enhance certification of security in real-world quantum communication.
We also note that the proposed method to calculate an effective error in protocols with side channels is intended to provide an intuitive tool for analyzing the influence of side channels in QKD. This method uses information flows to Bob and Eve to calculate an error value, capturing the possible eavesdropping of the side channel, which is unnoticeable through a standard communication error calculation. 

We demonstrated our ``effective error'' approach on an explicit joint eavesdropping strategy based on phase-covariant cloning of the operational degree of freedom followed by the joint collective measurement of a signal and a side channel degrees of freedom. 
As far as a phase-covariant cloning machine provides the most effective eavesdropping on the BB84 protocol without side channels, this is a reasonable example for the initial demonstration of our approach.
In our calculation results (dashed curves in Fig.~\ref{fig: Figure1} and Fig.~\ref{fig: Figure2}), the secret key rate drops to almost zero in 130 kilometers (eavesdropping on the side channel only, Fig.~\ref{fig: Figure1}) and in 100 kilometers (eavesdropping on both side channel and the signal state, Fig.~\ref{fig: Figure2}) while conservative estimates give an almost zero secret key rate at much shorter distances (30 kilometers in Fig.~\ref{fig: Figure1}, and 5 kilometers in Fig.~\ref{fig: Figure2} correspondingly). 
In our consideration, we investigated a non-adaptive eavesdropping strategy for the attack of the protocol with a passive source side channel. Based on our results, we conclude that if there is an attack that closes the gap between a conservative key rate estimate and a protocol without the side channel, then it must be an adaptive attack, i.e. an eavesdropping strategy that uses information from the side channel to change the operational degree of freedom in a favorable way to increase the information leakage. 

\section*{ACKNOWLEDGMENTS}

This work was supported by the Russian Science Foundation under grant no. 20-71-10072, https://rscf.ru/en/project/20-71-10072/.

\renewcommand{\appendixname}{APPENDIX}
\appendix


\section{Decoy state protocol with source side channel}

\subsection{BB84 protocol with decoy states}
\label{AppendixA1}

In practice, signal photons are generated by lasers with an unlocked initial phase, which leads to the randomization of the initial phases of signal photons. Furthermore, laser sources generate many-photon states along with single-photon states. The resulting carrier quantum state has the following form
\begin{equation}
 \rho^{\mu}_{x} = \sum_{k=0}^{\infty}e^{-\mu}\frac{\mu^{k}}{k!}\ket{k}\bra{k},
\end{equation}
where $\mu$ is the mean photon number of the coherent state. To decrease the influence of many-photon states on the communication the mean photon number is usually taken low (less than one photon per pulse), and the resulting state is called a phase-randomized weak coherent pulse (PRWCP).

Even though many-photon parts are attenuated, they continue to be a vulnerability in the BB84 protocol. These parts contain photons of the same single-photon states and thus open Eve a possibility to hold one photon with the same quantum state as photons sent towards Bob. In the end, Eve has a clone of the Bob state without introducing disturbance, and the security of the protocol is compromised.

This vulnerability can be closed with a so-called decoy-state method \cite{Ma2005}. 
This method is based on a fact, that even though Eve can measure a number of photons in a signal, she cannot deduce the signal intensity from such a measurement. Thus, if Alice and Bob use signals with different intensities, Eve will attack these signals with the same actions.
This fact allows Alice and Bob to use photon signals with different intensities to estimate the single-photon component weight, which composes the secret fraction of the distributed bits. They can further proceed with security amplification, taking into account only the secret fraction of their bits.

In particular, receiving $k$-photon states, where $k$ has values from 0 to infinity, the full probability of having a detection count on Bob side for a pulse with intensity is 
 \begin{equation}
     Q_{\mu} = \sum_{k=0}^{\infty}e^{-\mu}\frac{\mu^{k}}{k!}Y_{k} = Y_{0} + 1 - e^{-\eta \mu},
 \end{equation}
where $Y_{k}$ is a $k$-photon yield - a conditional probability that Bob detects a signal, when Alice sent him a $k$-photon state.
The probability of bit error on the Bob side is 
\begin{equation}
    E_{\mu} = \frac{1}{Q_{\mu}}\sum_{k=0}^{\infty}e^{-\mu}\frac{\mu^{k}}{k!}Y_{k}e_{k} = e_{0}Y_{0} + e_{det}(1 - e^{-\eta \mu}),
\end{equation}
where $k = 0$ corresponds to dark counts of photodetectors and $n$-photon error rate $e_{n}$ is 
\begin{equation}
    \label{en}
    e_{n} = \frac{e_{0}Y_{0} + e_{det}\eta_{n}}{Y_{n}}.
\end{equation}
Here $e_{0}$ is a probability of dark count, $Y_{0}$ is a vacuum yield, $Y_{n}$ is a $n$-photon yield, $e_{det}$ is an optical error of Bob's system. Here $\eta_{n}$ is the overall transmission and detection efficiency between Alice and Bob
\begin{equation}
    \eta_{n} = 1 - (1 - \eta)^{n},
\end{equation}
where 
\begin{equation}
    \eta = 10^{-\alpha L / 10}\eta_{Bob},
\end{equation}
where $\alpha$ is a loss coefficient, $L$ is a transmission line length and $\eta_{Bob}$ is a transmittance on the Bob's side. 
Decoy state protocol allows Alice and Bob to obtain bounds on a vacuum yield $Y_{0}$, a single-photon yield $Y_{1}$, and a single-photon error rate $e_{1}$ in a standard approach to BB84 protocol with decoy states. These quantities are further used to estimate the secret key rate. But, using (\ref{en}) allows to incorporate an explicit eavesdropping strategy, which Eve uses to attack the protocol. In the next section we will provide a connection between error from eavesdropping and a measurement error on Bob's side.

The secret key rate for the BB84 protocol with decoy states is
\begin{equation}
    \label{Rdecoy}
    R = \frac{1}{2}(Q_{1}(1 - h_{2}(e_{1})) - fQ_{\mu}h_{2}(E_{\mu}) ),
\end{equation}
where $f$ is a post-processing efficiency factor and $e_{1}$ is a bit error in single-photon component outcomes.

\subsection{Measurement errors and eavesdropping}
\label{AppendixA2}

Eavesdropping introduces errors in communication between Alice and Bob. In theory, they use perfect devices, and error occurs due to Eve interception. In practice, communication error arises from imperfections of devices. Mathematically, these two sources of errors (eavesdropping and measurement error) are equivalent, which we prove in this section. 

Suppose Alice sent $\ket{\Psi_{0}}\bra{\Psi_{0}}$ state in transmission channel and suppose that this state belongs to a basis \{$\ket{\Psi_{0}}\bra{\Psi_{0}}$, $\ket{\Psi_{1}}\bra{\Psi_{1}}$\}. In the channel, Eve uses a unitary attack, which entangles her subsystem with the Alice state:
\begin{equation}
    U(\ket{\Psi_{0}}\ket{E}) = \sqrt{1 - \eta}\ket{\Psi_{0}}\ket{0_{E}} + \sqrt{\eta}\ket{\Psi_{1}}\ket{1_{E}},
\end{equation}
where $\ket{E}$ - initial ancillary state, $\ket{0_{E}}$ and $\ket{1_{E}}$ ($\bra{1_{E}}\ket{0_{E}} = 0$) are two states of Eve subsystem. A unitary attack reads as a sequence of transforms: 
\begin{eqnarray}
    \notag
   \ket{\Psi_{0}}\bra{\Psi_{0}} \longrightarrow \ket{\Psi_{0}}\bra{\Psi_{0}} \otimes \ket{E}\bra{E}  \longrightarrow 
   (1 - \eta)\ket{\Psi_{0}}\bra{\Psi_{0}} \otimes \ket{0_{E}}\bra{0_{E}} + \\ \sqrt{\eta(1-\eta)}(\ket{\Psi_{0}}\bra{\Psi_{1}} \otimes \ket{0_{E}}\bra{1_{E}} + \ket{\Psi_{1}}\bra{\Psi_{0}} \otimes \ket{1_{E}}\bra{0_{E}}) + \eta \ket{\Psi_{1}}\bra{\Psi_{1}} \otimes \ket{1_{E}}\bra{1_{E}}.
\end{eqnarray}
The output state available to Bob is 
\begin{equation}
    \rho_{Bob} = (1 - \eta)\ket{\Psi_{0}}\bra{\Psi_{0}} + \eta \ket{\Psi_{1}}\bra{\Psi_{1}}.
\end{equation}
Bob measures this state in a particular basis, and the measurement is represented as a POVM operator $M$. Bob measures the expectation value of $M$ on a received quantum state, which contains eavesdropping-induced error. This state error can be transformed into measurement error:
\begin{eqnarray}
    \notag
   <M> = Tr[\rho_{Bob}M] = 
   Tr[((1 - \eta)\ket{\Psi_{0}}\bra{\Psi_{0}} + \eta \ket{\Psi_{1}}\bra{\Psi_{1}})M] = \\
   \notag
   (1 - \eta)Tr[\ket{\Psi_{0}}\bra{\Psi_{0}} M] + \eta Tr[\ket{\Psi_{1}}\bra{\Psi_{1}}M] = \\
   (1 - \eta)Tr[\ket{\Psi_{0}}\bra{\Psi_{0}} M] + \eta
   \notag
   Tr[V^{\dagger}\ket{\Psi_{0}}\bra{\Psi_{0}}V M] = \\
   \notag
   (1 - \eta)Tr[\ket{\Psi_{0}}\bra{\Psi_{0}} M] + \eta Tr[\ket{\Psi_{0}}\bra{\Psi_{0}} (VMV^{\dagger})] = \\
   Tr[\ket{\Psi_{0}}\bra{\Psi_{0}} ((1-\eta)M + \eta V^{\dagger}MV)] =
   Tr[\ket{\Psi_{0}}\bra{\Psi_{0}} \Tilde{M}],
\end{eqnarray}
where $V$ is a unitary transform $\ket{\Psi_{1}} = V \ket{\Psi_{0}}$. Here we denoted $\Tilde{M}$ an effective measurement device POVM, which incorporates a measurement error
\begin{equation}
    \Tilde{M} = (1 - \eta)M + \eta V^{\dagger}MV.
\end{equation}
This equivalence between state errors, which occur from eavesdropping, and measurement errors, which occur due to imperfect devices construction, allow describing eavesdropping as a optical measurement devices error on the Bob side and, formally, to add optical errors and theoretical communication error on the Bob side. Using (\ref{en}), we can incorporate theoretical error from eavesdropping into the single-photon error of decoy state method:
\begin{equation}
    \label{e1Q}
    e_{1} = \frac{e_{0}Y_{0} + (e_{det} + Q_{Bob})\eta}{Y_{1}}
\end{equation}
for a single-photon error and 
\begin{equation}
    \label{EmuQ}
    E_{\mu} = \frac{1}{Q_{\mu}}\sum_{k=0}^{\infty}e^{-\mu}\frac{\mu^{k}}{k!}Y_{k}e_{k} = e_{0}Y_{0} + (e_{det} + Q_{Bob})(1 - e^{-\eta \mu})
\end{equation}
for overall QBER in decoy state protocol.

\subsection{GLLP analysis of side channel}
\label{AppendixA3}

Influence of side channels on the QKD protocol is usually estimated with a GLLP-Koashi approach \cite{gottesman2002security, Koashi2009}. This approach introduces an additional quantum system - a quantum coin - to simulate Alice choice of basis. This quantum coin allows incorporating distinguishability between basis choices in the BB84 protocol into a quantity, which can be calculated once a model of side channel is provided. This approach to analyse information leakage of side channel is notoriously pessimistic and leads to a severe drop of QKD security even for practically weak side channels.

Here we apply this approach to a side channel model of our choice. Our derivation closely follows one of \cite{Lucamarini2015}. We introduced non orthogonal states as a model of side channel of a general kind (\ref{ensembleBB84sidechannel}). States of the ensemble are
\begin{eqnarray}
   \ket{\psi_{0X}} = \ket{0_{x}}\otimes\ket{0^{\Delta}_{X}}, \\
   \ket{\psi_{1X}} = \ket{1_{x}}\otimes\ket{1^{\Delta}_{X}}, \\
   \ket{\psi_{0Y}} = \ket{0_{y}}\otimes\ket{0^{\Delta}_{Y}}, \\
   \ket{\psi_{1Y}} = \ket{1_{y}}\otimes\ket{1^{\Delta}_{Y}}. \\
\end{eqnarray}
The X basis states can be prepared when Alice measures an ancillary qubit of the following entangled system
\begin{equation}
    \ket{\Psi_{X}} = \frac{
    \ket{0_{X}}\ket{\psi_{0X}} + \ket{1_{X}}\ket{\psi_{1X}}}
    {\sqrt{2}}
\end{equation}
and Y basis states can be prepared through measurement of the system
\begin{equation}
    \ket{\Psi_{Y}} = \frac{
    \ket{0_{Y}}\ket{\psi_{0Y}} + \ket{1_{Y}}\ket{\psi_{1Y}}}
    {\sqrt{2}}.
\end{equation}
To chose a basis, Alice can use a quantum coin system, which is entangled with two basis states
\begin{equation}
    \ket{\Phi} = \frac{\ket{0_{Z}}_{C} \ket{\Psi_{X}} + \ket{1_{Z}}_{C} \ket{\Psi_{Y}}}{\sqrt{2}},
\end{equation}
where subscript $C$ denotes a coin state. A full process of state preparation consists in measurement of a quantum coin to choose a basis and then measurement of a ancillary qubit to choose a secret bit.

In the absence of side channels, basis states $\ket{\Psi_{X}}$ and $\ket{\Psi_{Y}}$ are indistinguishable. The presence of side channels introduces a difference, which can be measured on a quantum coin. To see this, we rewrite states of quantum coin in $X$ basis 
\begin{equation}
    \ket{\Phi} = \frac{\ket{0_{X}}_{C} (\ket{\Psi_{X}}+\ket{\Psi_{Y}}) + \ket{1_{X}}_{C} (\ket{\Psi_{X}}-\ket{\Psi_{Y}}) }{2}.
\end{equation}
When basis states are indistinguishable ($\ket{\Psi_{X}} = \ket{\Psi_{Y}}$), the coin cannot be measured in a state $\ket{1_{X}}_{C}$, while when there is a distingusihability, there will be a non-zero probability to measure the coin in a state $\ket{1_{X}}_{C}$.
This probability is equal to
\begin{equation}
    \label{Delta}
    \Delta = Pr(\ket{1_{X}}_{C}) = \frac{1 - Re[\bra{\Psi_{X}}\ket{\Psi_{Y}}]}{2}.
\end{equation}

If we substitute side channel state from model of our choice, we will obtain a following form
\begin{equation}
    \label{Delta}
    \Delta = \frac{
    4 - 
    \bra{0^{\Delta}_{X}}\ket{0^{\Delta}_{Y}} -
    \bra{0^{\Delta}_{X}}\ket{1^{\Delta}_{Y}} - 
    \bra{1^{\Delta}_{X}}\ket{0^{\Delta}_{Y}} -
    \bra{1^{\Delta}_{X}}\ket{1^{\Delta}_{Y}} 
    }{8}.
\end{equation}
In case, when all side channel states have equal inner products ($\bra{i^{\Delta}_{X}}\ket{j^{\Delta}_{Y}} = S, i \neq j$), this form simplifies to 
\begin{equation}
    \Delta = \frac{1 - S}{2}.
\end{equation}
The form (\ref{Delta}) allows us to compare our approach for estimating side channel influence with an approach of GLLP-Koashi. 

This value allows to calculate a new error value, which takes into account side channel information leakage. For a decoy-state protocol, this error values reads
\begin{equation}
    \label{e1prime}
    e_{1}^{'} = e_{1} + 4(1 - \Delta^{'})\Delta^{'}(1 - 2e_{1}) + 4(1 - 2\Delta^{'})\sqrt{\Delta^{'}(1 - \Delta^{'})e_{1}(1 - e_{1})},
\end{equation}
where 
\begin{equation}
    \Delta^{'} = \frac{\Delta}{Y_{1}}
\end{equation}
is a corrected value, which takes into account the ability of Eve to use lossless channel. The secret key rate with side channel under the GLLP-Koashi analysis is 
\begin{equation}
    \label{RGLLP}
    R = \frac{1}{2}(Q_{1}(1 - h_{2}(e_{1}^{'})) - fQ_{\mu}h_{2}(E_{\mu}) ).
\end{equation}

 
\bibliographystyle{apsrev4-1}
\bibliography{ref}

\end{document}